\documentstyle[twoside,slashed,fleqn,espcrc2,epsf]{article}

\newcommand{\half}{\mbox{\small $\frac{1}{2}$}}          

\def\lsim{\mathrel{\rlap{\lower4pt\hbox{\hskip1pt$\sim$}}
    \raise1pt\hbox{$<$}}}                
\def\gsim{\mathrel{\rlap{\lower4pt\hbox{\hskip1pt$\sim$}}
    \raise1pt\hbox{$>$}}}                
\def\ring{\mathaccent"7017}

\title{
       Determining $g_A$ using non-perturbatively $O(a)$
       improved Wilson fermions }

\author{
        R.~Horsley%
           \address{Deutsches Elektronen-Synchrotron DESY \& NIC,
                    D-15735 Zeuthen, Germany}
                       \hspace{-0.25cm} $^,$\hspace{-0.15cm}
           \address{Institut f\"ur Physik, Humboldt-Universit\"at zu Berlin,
                    D-10115 Berlin, Germany},
           {\it QCDSF} and {\it UKQCD} Collaborations }
       
\begin{document}

\begin{abstract}
A completely non-perturbative estimate is given for $g_A$
using both quenched and unquenched $O(a)$ improved
Wilson fermions. Particular attention is paid to
the determination of the axial renormalisation
constant, $Z_A$, using the Ward identity for the propagator.
For the quenched case, we have results at three lattice
spacings allowing a continuum extrapolation.
\end{abstract}

\maketitle

\setcounter{footnote}{0}


\section{INTRODUCTION}
\label{introduction}

The axial charge of the nucleon, $\Delta q$, is defined from the axial
current by
\begin{eqnarray}
   \langle N(\vec{p}, \vec{s})| {\cal A}^\mu_R | N(\vec{p}, \vec{s}) \rangle
               = 2 s^\mu \Delta q ,
                                               \nonumber
\end{eqnarray}
with ${\cal A}_R^\mu = (\overline{q}\gamma^\mu\gamma_5 q)_R$,
giving for the non-singlet charge
\begin{eqnarray}
   g_A \equiv g_A^{(3)} = \Delta u - \Delta d .
                                               \nonumber
\end{eqnarray}
This determines the strength of $\beta$-decay and also occurs
in the Bjorken sum rule where the lowest moment of the
difference between the proton and neutron
polarised structure function $g_1$ is proportional to the non-singlet
charge. To find $g_A$ the programme is:
\begin{enumerate}
   \item Compute $\langle N| \overline{u}\gamma_\mu\gamma_5 u -
         \overline{d}\gamma_\mu\gamma_5 d |N\rangle$
         on the lattice for {\it many} quark masses ($\gg 3$) and
         {\it many} lattice spacings ($\gg3$).
         \label{lattice_comp}
   \item Determine non-perturbatively the renormalisation constant $Z_A$.
         \label{np_za}
   \item Chirally extrapolate $m_q$ (or $m_\pi^2$) $\to 0$.
         \label{chiral_extrap}
   \item Continuum extrapolate $a^2 \to 0$
         (if using $O(a)$ improved fermions)
   \item Do first for quenched (as much cheaper in {\it CPU} time)
         then repeat for unquenched fermions.
   \item Compare with experiment, $g_A \approx 1.26$.
\end{enumerate}
Our lattice simulation, point (\ref{lattice_comp}), is standard using ratios
of nucleon three-point to two-point correlation functions and will
not be discussed further here, except to note that because we are
considering a non-singlet function, the difficult to compute
quark line disconnected terms cancel. Here, in this talk,
we shall mainly discuss point (\ref{np_za}).


\section{DETERMINING $Z_A$}

On the lattice, for Wilson fermions, the chiral Ward Identity is
only approximately true, due to discretisation effects. 
We have a {\it finite} renormalisation constant $Z_A(g)$ to determine
by demanding that a Ward Identity is obeyed to $O(a^n)$
(with $n=2$ for $O(a)$ improved fermions). The ALPHA collaboration
\cite{luscher96a} has found $Z_A$ for quenched fermions using
the Schr\"odinger functional and {\it PCAC}; here, alternatively,
we shall use the chiral Ward identity for the propagator.
We shall first check our $Z_A$ results for quenched fermions
before then determining $Z_A$ for unquenched fermions.
The chiral Ward identity for the propagator, $S_R$, reads
\begin{eqnarray}
   \gamma_5 S_R^{-1} + S_R^{-1}\gamma_5 = 2 \; m_R \; \Lambda_R^P ,
                                               \nonumber
\end{eqnarray}
where $m_R$ is the quark mass,
and $\Lambda_R^P$ the $1PI$ pseudoscalar vertex,
obtained from the Green's function $G_P$ by
amputating the external propagators.
On the lattice we take this equation
to be correct to $O(a^2)$,
\begin{eqnarray}
   \mbox{Tr} S_*^{-1} = Z_A \widetilde{m} \mbox{Tr} \gamma_5 \Lambda_*^P ,
                                               \nonumber
\end{eqnarray}
where $O(a)$ improved operators are denoted with a star.
The bare quark mass $m_*$ has also been replaced by the quark mass
obtained%
\footnote{A small additional factor $1 - am(b_P-b_A)$ has been absorbed
into a redefined $\Lambda_*^P$.}
from {\it PCAC}, $\widetilde{m}$.
These quark masses may be computed in the
standard way using the appropriate axial vector matrix elements.
The propagator and vertex can also be found
on the lattice, in the Landau gauge, \cite{martinelli94a,gockeler98a}.
A problem with using the propagator is that we now need $O(a)$
improvement {\it off-shell}. To try to ameliorate this problem
we note that as dominant {\it off-shell} effects come at short
distances, we shall subtract a {\it contact term} (ie delta function
in position space), \cite{heatlie91a}, from the propagator
and Green's function in addition to improving the appropriate
operator.
This procedure has been shown to work
to first order in perturbation theory, \cite{capitani00a},
and we shall use the formulation given there. Note however
that the pseudoscalar Green's function (or $1PI$ vertex)
is particularly simple: there is effectively only one additional $O(a)$
improvement term. However, initial attempts
to satisfy the above equation fail because of the
presence of large $O(ap^2)$ terms mainly due to the Wilson term
in the propagator, and so here we choose to re-write the propagator $S_*$
and vertex $\Lambda_*^P$ in an $O(a)$ equivalent form, \cite{capitani97a},
as solutions of
\begin{eqnarray}
   S^{-1} = \left( 1 - am b_\psi {\widehat{p}^2\over \ring{p}^2}
                     \right) S^{-1}_* 
                     - \half a \lambda_\psi {\widehat{p}^2\over \ring{p}^2}
                       S^{-2}_*,
                                               \nonumber
\end{eqnarray}
and
\begin{eqnarray}
   \left[ 1 - am d_P \right] \Lambda_*^P 
          = \Lambda^P + \half a\lambda_\psi \{ S_*^{-1}, \Lambda_*^P \},
                                               \nonumber
\end{eqnarray}
($\ring{p}_\mu = (1/a) \sin ap_\mu$, $\widehat{p}_\mu = (2/a) \sin ap_\mu/2$)
where $b_\psi = -1 + O(g^2)$ and $\lambda_\psi = 1 + O(g^2)$
are the improvement coefficient and coefficient of the
contact term for the propagator and $d_P = b_\psi - \lambda_P Z_mZ_P + b_A
= -1 + O(g^2)$.
This gives $S_*^{-1} = i\slashed{\ring{p}}+m_* + O(a^3m^2p^2)$
for the free field. The Wilson term has been suppressed,
$O(ap^2)$ terms being multiplied by an additional small $a^2m^2$
term. As in this case
$\mbox{Tr} \gamma_5 \Lambda_*^P \sim 1/(1 + \half (am)^2)$
it is little effected from its true value of $1$.

\begin{figure}[tb]
   \epsfxsize=7.00cm \epsfbox{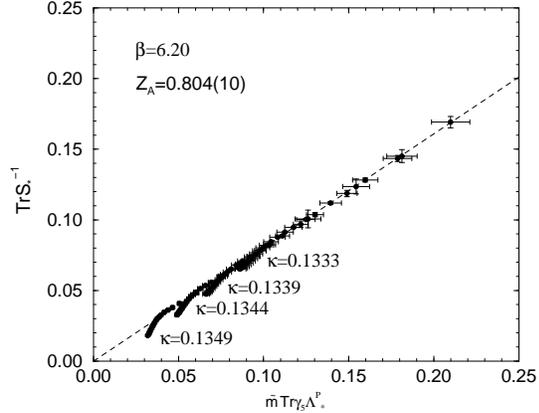}  
   \vspace*{-0.25in}
   \caption{\footnotesize{\it $\mbox{Tr} S_*^{-1}$ plotted against
            $\widetilde{m}\mbox{Tr}\gamma_5 \Lambda_*^P$
            for $\beta = 6.20$ for $4$ quark masses.
            Heavier masses lie on the RHS of the
            picture. Smaller momenta are on the RHS
            of each data set.}}
   \vspace*{-0.25in}
   \label{fig_rat_b6p20_lat00}
\end{figure}

\begin{figure}[tb]
   \epsfxsize=7.00cm \epsfbox{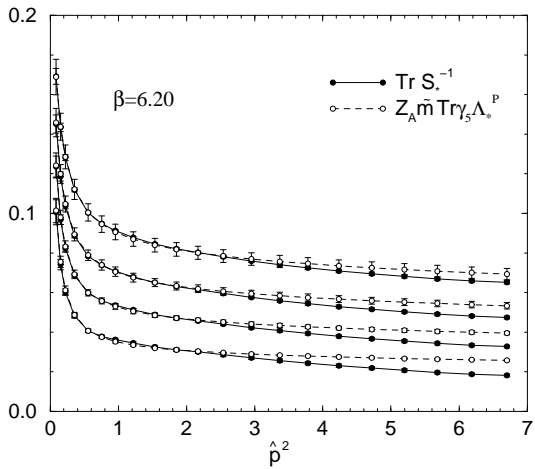}  
   \vspace*{-0.25in}
   \caption{\footnotesize{\it $\mbox{Tr} S_*^{-1}$ plotted with
            $Z_A \widetilde{m}\mbox{Tr}\gamma_5 \Lambda_*^P$
            for $\beta=6.20$ for the $4$
            quark masses (the heavier quark masses are higher).}}
   \vspace*{-0.25in}
   \label{fig_trSm1+trLam_b6p20_lat00} 
\end{figure}

\begin{figure}[tb]
   \epsfxsize=7.00cm \epsfbox{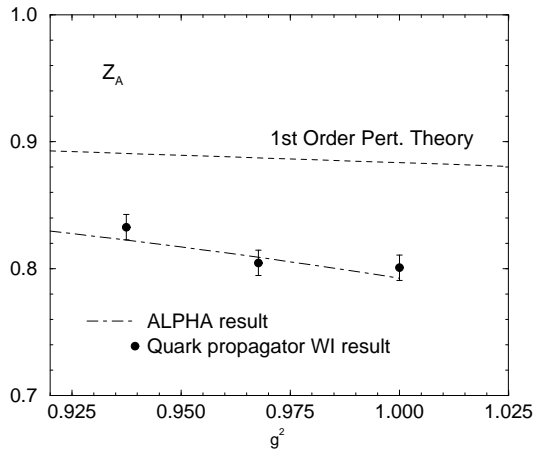}  
   \vspace*{-0.25in}
   \caption{\footnotesize{\it $Z_A$ versus $g^2$ (filled circles).
            Also shown is the ALPHA non-perturbative determination,
            \cite{luscher96a}, together with
            the first order perturbation theory result.}}
   \vspace*{-0.25in}
   \label{fig_za_g2_magic_lat00}
\end{figure}

We shall first determine $Z_A$ for quenched fermions.
We have generated data sets at $\beta = 6.0$, $6.2$, $6.4$,
each data set with $3$ (or more) quark masses.
To determine $Z_A$, we attempt to minimise
$R(p) \equiv \mbox{Tr}S_*^{-1}/\widetilde{m}\mbox{Tr}\gamma_5\Lambda_*^P 
= \mbox{const}$ for many momenta (a three parameter fit). 
A typical result is given in Fig.~\ref{fig_rat_b6p20_lat00}.
The gradient gives an estimate for $Z_A$. Deviations are seen
for the higher momenta. This is more clearly seen in 
Fig.~\ref{fig_trSm1+trLam_b6p20_lat00} 
which tells us how far we can go, a good fit being obtained
for $(a\widehat{p})^2 \lsim 2.5$, say.
Once the quark propagator has been obtained, we may then use it 
to find $Z^{MOM}_\psi$ and $m_q^{MOM}$, \cite{becirevic99a}.
We shall not be concerned with this here, \cite{gockeler00a},
but just concentrate on $Z_A$. In Fig.~\ref{fig_za_g2_magic_lat00}
we compare our result here with the previously known
result, \cite{luscher96a}. Reasonably good agreement is
found. (In any case, as different methods have different $O(a^2)$ corrections
complete agreement does not have to be found.)

\begin{figure}[tb]
   \epsfxsize=7.00cm \epsfbox{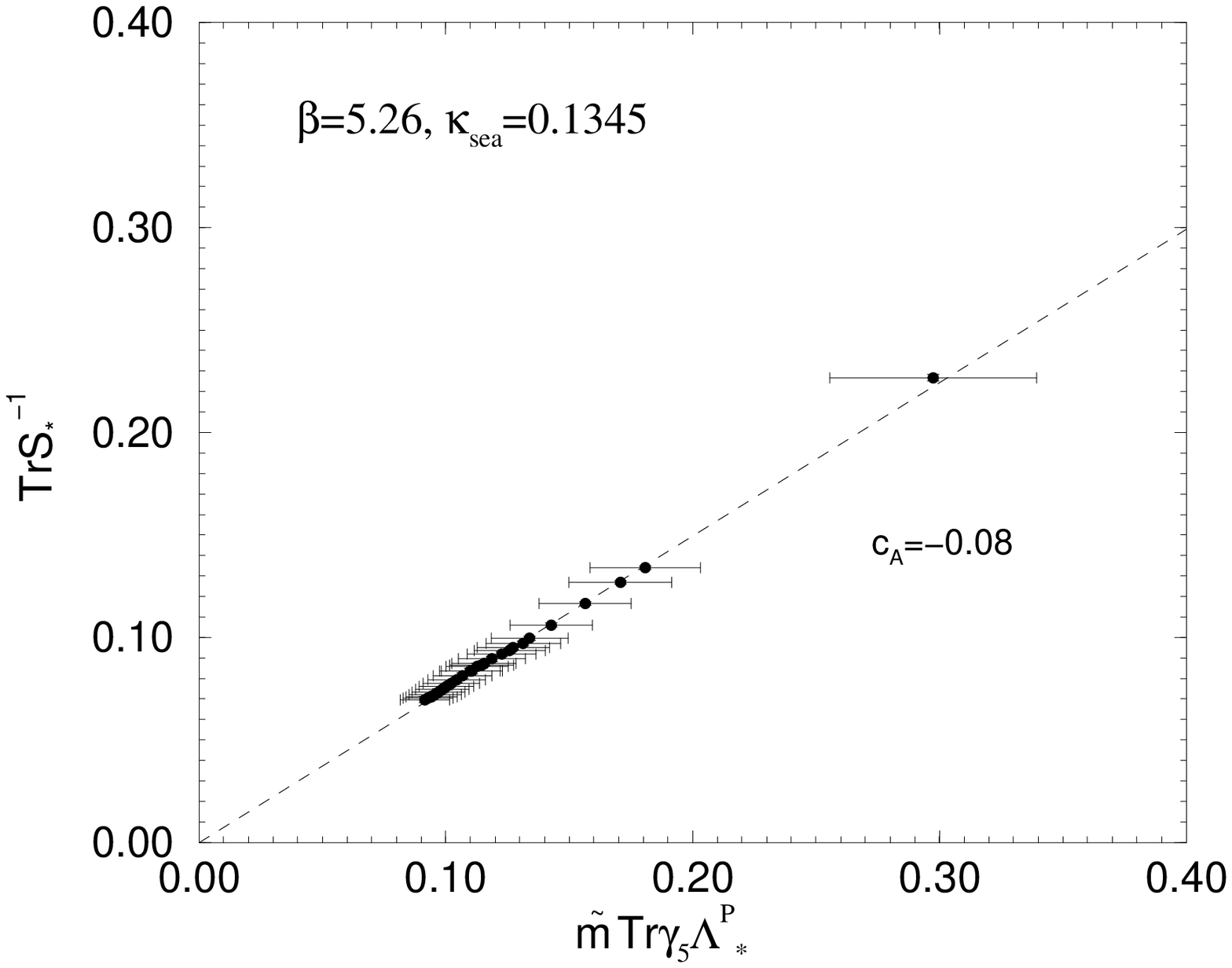}
   \vspace*{-0.25in}
   \caption{\footnotesize{\it $\mbox{Tr} S_*^{-1}$ plotted against
            $\widetilde{m}\mbox{Tr}\gamma_5 \Lambda_*^P$. }}
   \vspace*{-0.25in}
   \label{fig_rat_b5p26kp1345c1p9497_lat00}
\end{figure}

Buoyed up by this we now turn to unquenched $n_f = 2$ 
fermions where we have analysed three {\it UKQCD} data sets, \cite{ukqcd},
$(\beta, \kappa_{sea}) =$ (5.29,0.1340), (5.26,0.1345) and
(5.20,0.1350). These are {\it matched} at an approximately
constant $r_0$ value, corresponding to $a \sim 0.105\mbox{fm}$
(using $r_0 = 0.5\mbox{fm}$).
A typical result is shown in Fig.~\ref{fig_rat_b5p26kp1345c1p9497_lat00}.
The {\it preliminary} values
obtained are given in Table~\ref{table_dynamica_za}.
\begin{table}[tb]
\begin{center}
   \begin{tabular}{||l|l|l||}
      \hline
      $(\beta,\kappa_{sea})$ & $Z_A|_{c_A=-0.02}$
                                          & $Z_A|_{c_A=-0.08}$  \\
      \hline
      $(5.29, 0.1340)$   & 0.65(2)  & 0.73(2)  \\
      $(5.26, 0.1345)$   & 0.68(3)  & 0.75(2)  \\  
      $(5.20, 0.1350)$   & 0.64(1)  & 0.71(1)  \\
      \hline
   \end{tabular}
\end{center}
\caption{\footnotesize{\it Preliminary values of $Z_A$ obtained for
         unquenched fermions.}}
\label{table_dynamica_za}
\vspace*{-0.25in}
\end{table}
Unfortunately for constructing $\widetilde{m}$, the
axial current improvement coefficient $c_A$ is unknown;
we shall use here two values, the first $c_A \sim -0.02$ corresponding
to a tadpole improved result, while the second $c_A \sim -0.08$
is the non-perturbative quenched value at $\beta =6.0$.
While this does not change the fit, ie the product $Z_A\widetilde{m}$,
the $10\%$ change in $\widetilde{m}$ directly produces a $10\%$
change in $Z_A$.

\begin{figure}[tb]
   \epsfxsize=7.00cm \epsfbox{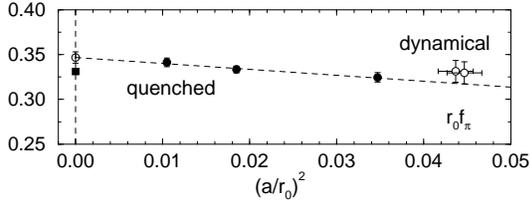}
   \vspace*{-0.25in}
   \caption{\footnotesize{\it $r_0f_\pi$ versus $(a/r_0)^2$, with 
            the quenched results (filled circles, the continuum
            extrapolated value being given by an open circle),
            together with the unquenched results, (open circles).
            Note here that because $f^{bare}_\pi$ decreases as $c_A$
            decreases while $Z_A$ increases, changes in $c_A$
            have little effect on the result.
            The experimental value is shown as a filled square.}}
   \vspace*{-0.25in}
   \label{fig_fpi_quen+dyn_lat00}
\end{figure}

As a further check we have first computed $f_\pi$.
After performing the chiral extrapolation, for the three
data points given in Table~\ref{table_dynamica_za}, we find
the results given in Fig.~\ref{fig_fpi_quen+dyn_lat00}.
Although a satisfactory result is obtained,
one should not read too much into this at present,
as apart from the uncertainty in $Z_A$
only three quark masses are used in the chiral extrapolation.


\section{RESULTS FOR $g_A$}

\begin{figure}[tb]
   \epsfxsize=7.00cm \epsfbox{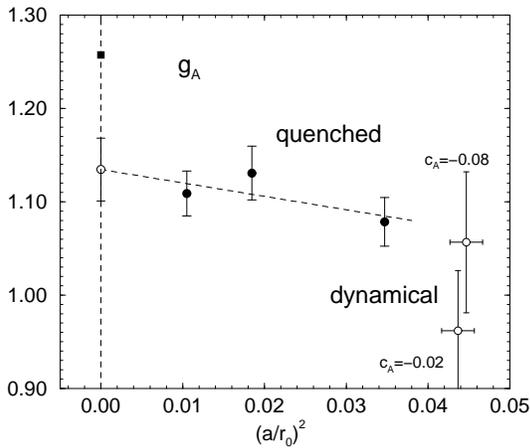}
   \vspace*{-0.25in}
   \caption{\footnotesize{\it $g_A$ versus $(a/r_0)^2$, with
            the quenched results (filled circles, the continuum
            extrapolated value being given by an open circle),
            together with the unquenched results, (open circles).
            The experimental value is shown as a filled square.}}
   \vspace*{-0.25in}
   \label{fig_ga_aor02_p0+dyn_lat00}
\end{figure}

We are now ready to present our results for $g_A$. Again after
the (linear) chiral extrapolation%
\footnote{A typical chiral extrapolation for $\beta=6.0$
quenched fermions is shown in Fig.~1 of \cite{capitani99a}.
(Note that we take the unknown improvement coefficient $b_A = 0$ 
here. We have checked that in the chiral limit, as expected,
this plays no role in the numerical value.)}
we plot our results as a function of $a^2$.
The current results are shown in Fig.~\ref{fig_ga_aor02_p0+dyn_lat00}.
All results, at present, lie too low in comparison with the
experimental number. Even allowing for the uncertainty
in the determination of $Z_A$ and hence $g_A$ for unquenched fermions,
it would still seem that the result is around the quenched result.
(Note that \cite{gusken99a} also find a low value of $g_A$.)
Further details will be published elsewhere, \cite{gockeler00b}.


\section{CONCLUSIONS}

A completely non-perturbative evaluation of a nucleon matrix element needs
\begin{itemize}
   \item large physical boxes
   \item good determination of $Z$
   \item chiral extrapolation (delicate)
   \item continuum extrapolation (delicate) 
\end{itemize}
How far have we got? We have concentrated on $g_A$ here
and tried to develop a way of determining $Z_A$ using the propagator
and pseudoscalar vertex. It is apparent though that we are only at the
beginning, not only for the unquenched results, but even for
the quenched results many more quark mass and $a$ values must
be used to allow for reasonably reliable chiral and continuum extrapolations.
We finally note that the successful determination of $g_A$ 
is a real test of {\it QCD} as it is a relatively
simple (nucleon) matrix element, well measured in experiment.


\section*{ACKNOWLEDGEMENTS}

The numerical computations were performed on the
Quadrics machines at NIC (Zeuthen) as well as the
Cray {\it T3E}s at ZIB (Berlin), EPCC (Edinburgh) and
NIC (J\"ulich). We wish to thank all these institutions for their support.
UKQCD acknowledges PPARC grants GR/L22744 and
PPA/G/S/1998/000777.


\end{document}